\documentclass[twocolumn,prb]{revtex4}
\usepackage{graphicx}
\usepackage{dcolumn}
\usepackage{bm}
\usepackage{amsmath}
\usepackage{times}
\usepackage{color}
\usepackage[breaklinks=true,colorlinks,citecolor=blue,linkcolor=blue,urlcolor=blue]{hyperref}

\makeatletter

\newcommand{\Rmnum}[1]{\expandafter\@slowromancap\romannumeral #1@}
\makeatother

\begin{document}

\author{Bikash Ghosh}
\affiliation{Department of Physics, Indian Institute of Technology Kanpur, Kanpur 208016, India}
\author{Soumik Mukhopadhyay}
\email{soumikm@iitk.ac.in}
\affiliation{Department of Physics, Indian Institute of Technology Kanpur, Kanpur 208016, India}
\title{Unified description of resistivity and thermopower of Pr$_{2}$Ir$_{2}$O$_{7}$ : possible influence of crystal field excitation in a Kondo lattice}
\begin{abstract}
We present experimental evidence of incoherent Kondo scattering as the source of resistivity minima in bulk polycrystalline and nanocrystalline Pr$_{2}$Ir$_{2}$O$_{7}$. The temperature dependence of thermopower shows a positive maximum at high temperature followed by a negative minimum at low temperature, with the sign inversion occurring at a much higher temperature than $T_K$. Moreover, we observe little correlation between $T_K$ and inter-site coupling strength given by $|\theta_{CW}|$. We describe the temperature dependence of thermopower and resistivity within the framework of crystal field excitation in a Kondo lattice.
\end{abstract}

\maketitle

Among the pyrochlore iridates, R$_{2}$Ir$_{2}$O$_{7}$ ($R=Y$ or lanthanide element), resistivity enhancement with decreasing temperature is ubiquitous and the degree of resistivity enhancement depends, \textit{inter alia}, on the R site atomic radius~\cite{Yanagishima,Krempa,Matsuhira,Takatsu,Abhishek1,Abhishek2}. However, the marginal resistivity upturn in the metallic pyrochlore iridate Pr$_{2}$Ir$_{2}$O$_{7}$ (PIO)~\cite{Nakatsuji,Ohtsuki,Cheng,Machida} has been one of the most contentious topics over the past decade or more. PIO is a metallic spin liquid with novel topological properties at low temperature \cite{Machida, Machidaa,Foronda,Onoda}. The rare-earth 4f Pr$^{3+}$ local magnetic moments interact with each other through RKKY interaction mediated by 5d Ir$^{4+}$ conduction electrons~\cite{Lee,Ikeda}. Non-coplanar spin textures of both $Pr^{3+}$ and $Ir^{4+}$ spins lead to unconventional magneto-transport properties including spontaneous hysteretic hall effect without magnetic long range order at low temperature~\cite{Machida,Machidaa,Balicas, Machida1}. However, till date, the origin of low temperature resistivity minimum in PIO is not fully understood~\cite{Nakatsuji,Cheng,Udagawa}. In PIO, $Pr^{3+}$ has a magnetic non-Kramers doublet ground state, residing on the vertex of corner sharing tetrahedra with the Ising axes along [111] directions~\cite{Gardner,Krempa}. This makes it a candidate two-channel Kondo system if valence fluctuation to the Kramers doublet excited state is taken into account. However, previous experiments seem to suggest that the resistivity upturn arises from `inter-site' Kondo coupling, much like `two impurity' single channel Kondo scattering problem, largely dominated by RKKY interaction between $Pr^{3+}$ local moments~\cite{Nakatsuji, Kimura}.

Recently, the angle resolved photoemission spectroscopy (ARPES) measurement showed that PIO possesses $3$D quadratic band touching (QBT) at Brillouin zone center~\cite{Kondo}. The QBT systems are generally strongly interacting~\cite{Abrikosov}. Thus doping, strain or confinement effects could potentially lead to formation of a large variety of strongly correlated quantum phases such as topological Mott insulator, magnetic Weyl semimetal, and anomalous Hall states~\cite{Kondo}. The low frequency electrodynamic response of bi-axially compressed PIO films suggests that the resistivity minimum is a consequence of interplay between decreasing scattering rate as well as carrier density in a slightly doped system~\cite{Cheng}. However, it is unclear whether such a mechanism should be applicable in general. CDMFT calculations suggest a different origin of the resistivity minimum: scattering of conduction electrons by `two-in, two-out' spin texture~\cite{Udagawa}.

\begin{figure}
\includegraphics[width=\linewidth]{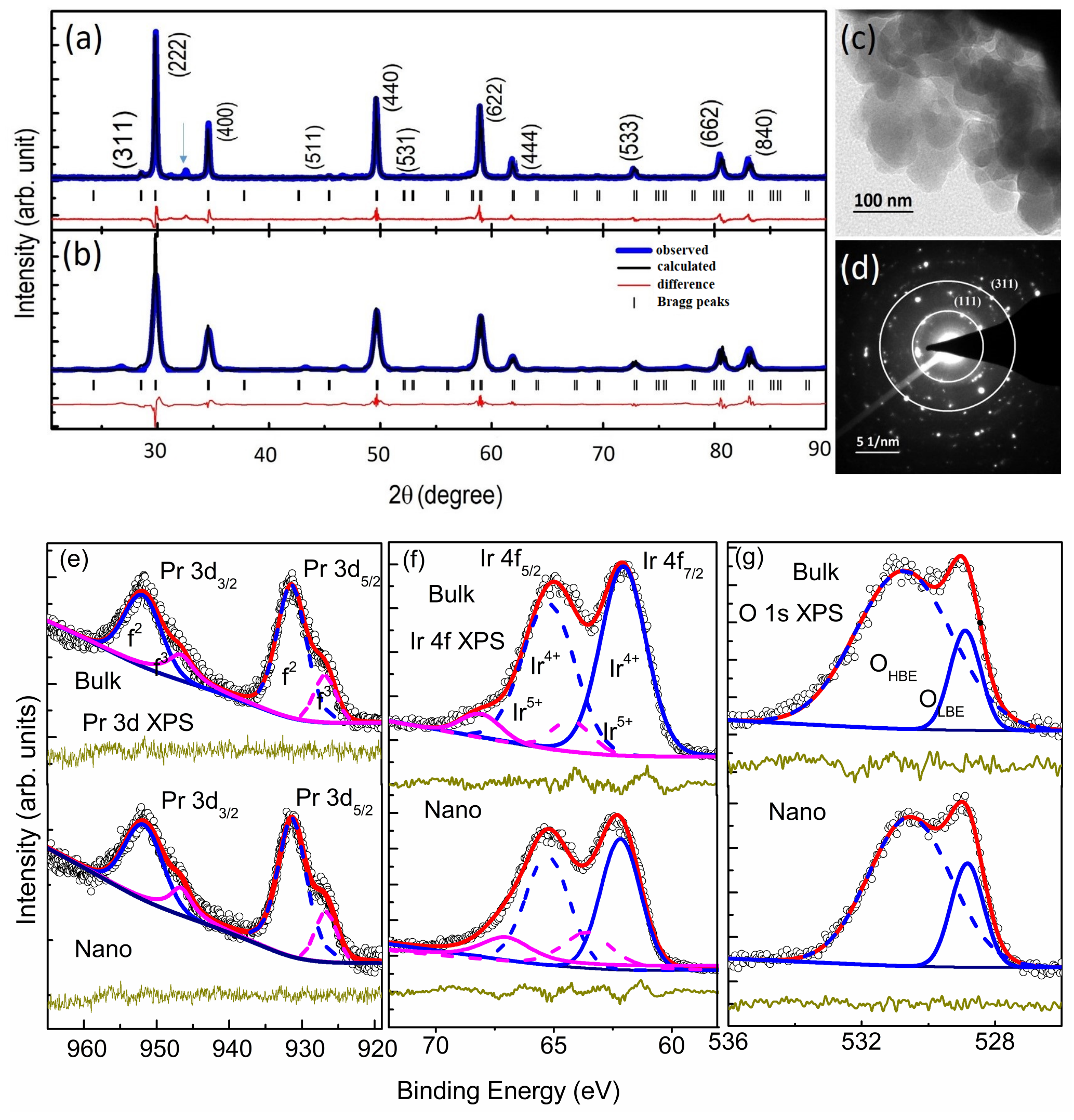}
\caption{Room temperature X-ray diffraction pattern of (a) bulk polycrystalline and (b) nanocrystalline PIO. The experimental and Reitveld refinement data are shown. The position of Bragg reflections are shown by vertical bars. The difference between experimental and refined data is plotted by continuous (red) lines at the bottom. The nominal presence of secondary phase Pr$_6$O$_{11}$ in bulk PIO is shown by the arrow in a) while no such extra phase is observed in nanocrystalline PIO. (c) Transmission electron micrograph (TEM) image shows average particle size of $50$ nm in nanocrystalline PIO. (d) Selected Area Electron Diffraction (SAED) for the nanocrystalline sample. (e)-(g) X-ray photoelectron spectra (XPS) for bulk and nanocrystalline PIO: Pr - 3d, de-convoluted Ir 4f spectra, and O-1s spectra.}\label{fig1}
\end{figure}

The key finding of the present article is the temperature dependence of thermoelectric power in Pr$_2$Ir$_2$O$_7$. For both bulk polycrystalline as well as nanocrystalline samples, the thermopower exhibits a shallow positive maximum at high temperature and a prominent negative minimum at low temperature. The temperature dependence of thermopower is analyzed using a phenomenological model incorporating crystal field excitation in a Kondo lattice. The same phenomenological model with unchanged crystal field parameters is subsequently used to describe the observed low temperature resistivity minimum.

Bulk poly-crystalline PIO was grown by conventional solid state reaction method~\cite{Yanagishima,Dwivedi}. Nanocrystalline PIO sample was synthesized using chemical sol-gel route and subsequent sintering~\cite{Dwivedi}. The starting materials used were, Praseodymium Oxide Pr$_{6}$O$_{11}$(Sigma-Aldrich, 99.99$\%$) and Iridium Acetate IrC$_{6}$H$_{9}$O$_{6}$ [Alfa-Aesar, Ir (48-54\%)]. X-ray diffraction (XRD) measurement was carried out at room temperature using PANalytical XPertPro diffractometer. Particle size of the chemically synthesized nanocrystalline sample was verified independently by Field emission scanning electron microscopy (FESEM) (TITAN $G^{2}$) and XRD. The oxidation states were identified by using X-ray photoelectron spectroscopy (XPS) using PHI $5000$ Versa Probe II system. Electrical transport and thermopower measurements were carried out using Quantum design physical property measurement system (PPMS) and the magnetic susceptibilities were measured by Quantum design SQUID magnetometer.

X-ray diffraction patterns of bulk polycrystalline and nanocrystalline PIO along with structural Rietveld refinement are shown in Fig.~\ref{fig1}(a) and (b), respectively. We used FULLPROF software for Rietveld refinement \cite{Reitveld}. The atomic coordinates of Pr, Ir , O1 and O2 for bulk polycrystalline PIO are similar to already reported values~\cite{Takatsu} and for nanocrystalline PIO the same are 16d (1/2, 1/2,1/2), 16c (0,0,0), 48f (0.328 ($x_{o1}$), 1/8,1/8) and 8b (3/8,3/8,3/8), respectively. The $x$ coordinate of O1 is slightly higher in bulk polycrystalline ($x_{o1}$=0.330) PIO compared to nanocrystalline samples. The lattice parameters obtained from the refinement are $a=10.42\AA$, $V=1131.36\AA^{3}$ ($\chi^{2}=3.46$) for bulk and $a= 10.40\AA$, $V=1124.86\AA^{3}$ ($\chi^{2}=4.26$) for nanocrystalline PIO, respectively. The average crystallite size for nanocrystalline PIO calculated from XRD using Debye Scherer formula turns out to be $50$ nm. We also calculated average particle size form FE-SEM image (Fig.~\ref{fig1}(c)) using ImageJ software, which is in fair agreement with the result obtained from XRD. The selected area diffraction (SAED) pattern shows diffraction rings of the nano-crystalline sample (Fig.~\ref{fig1}(d)).

\begin{figure}
\includegraphics[width=\linewidth]{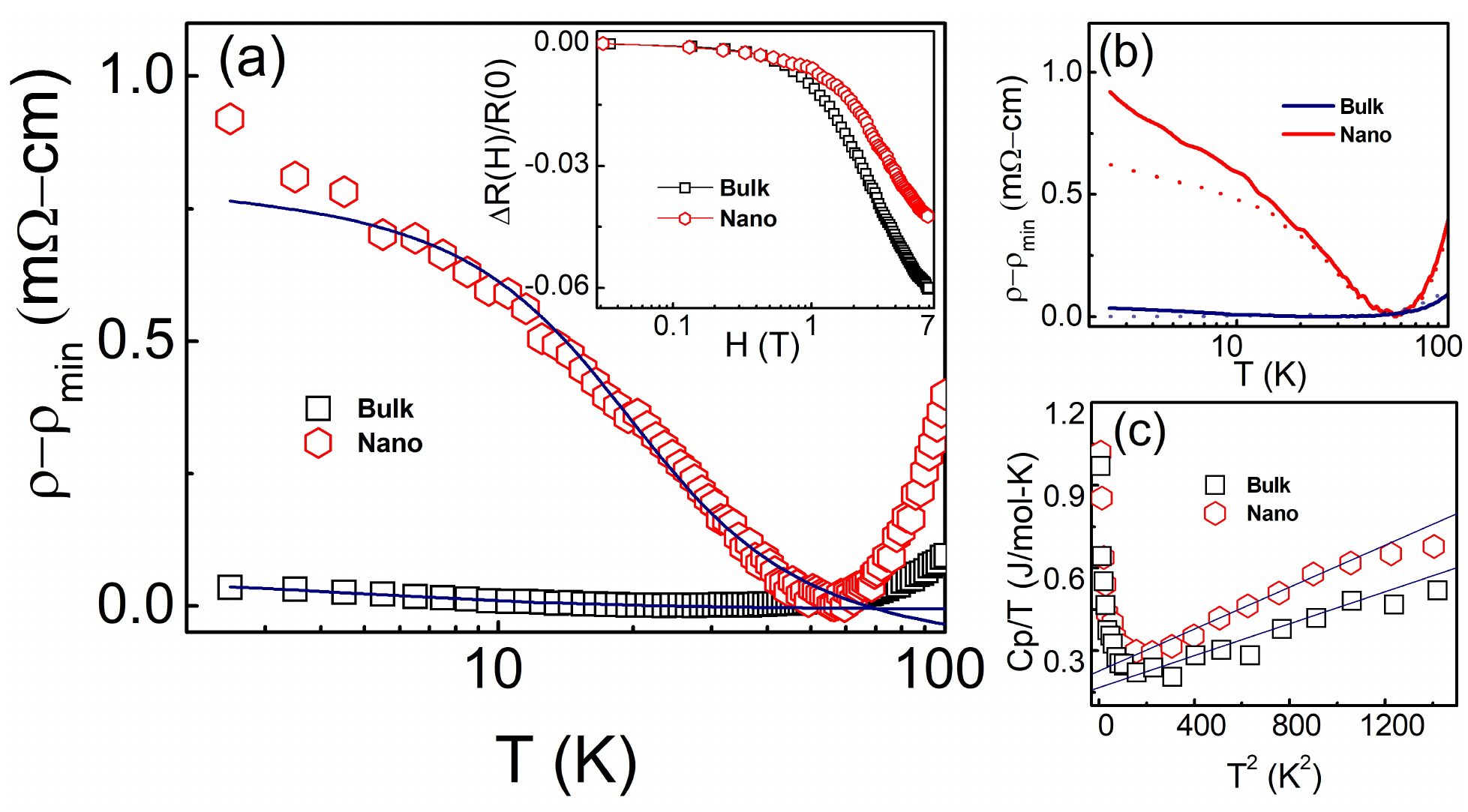}
\caption{(a) Spin contribution to the resistivity($\rho_{m}=\rho(T)-\rho_{min}$) as a function of $T$ for bulk and nanocrystalline PIO. The minima in resistivity ($\rho_{min}$) were observed at $27$~K and $55$~K for bulk and nanocrystalline PIO, respectively. The spin contribution is much higher in nanocrystalline sample compared to the bulk. The solid lines are the theoretical fits. Inset: Magnetic field dependence of magnetoresistivity at $5$ K. (b) The temperature dependence of resistivity in absence and in presence of constant magnetic field $5$ T applied perpendicular to the current shows partial suppression of resistivity upturn around $T_K$, leading to negative magnetoresistance. (c) Specific heat divided by temperature $C_{P}$/T is plotted against $T^{2}$ for bulk and nanocrystalline PIO. The solid line passing through the data points shows linear fit with $C_{p}/T = \gamma+\beta T^{2}$.}\label{fig2}
\end{figure}
\begin{figure}
\includegraphics[width=\linewidth]{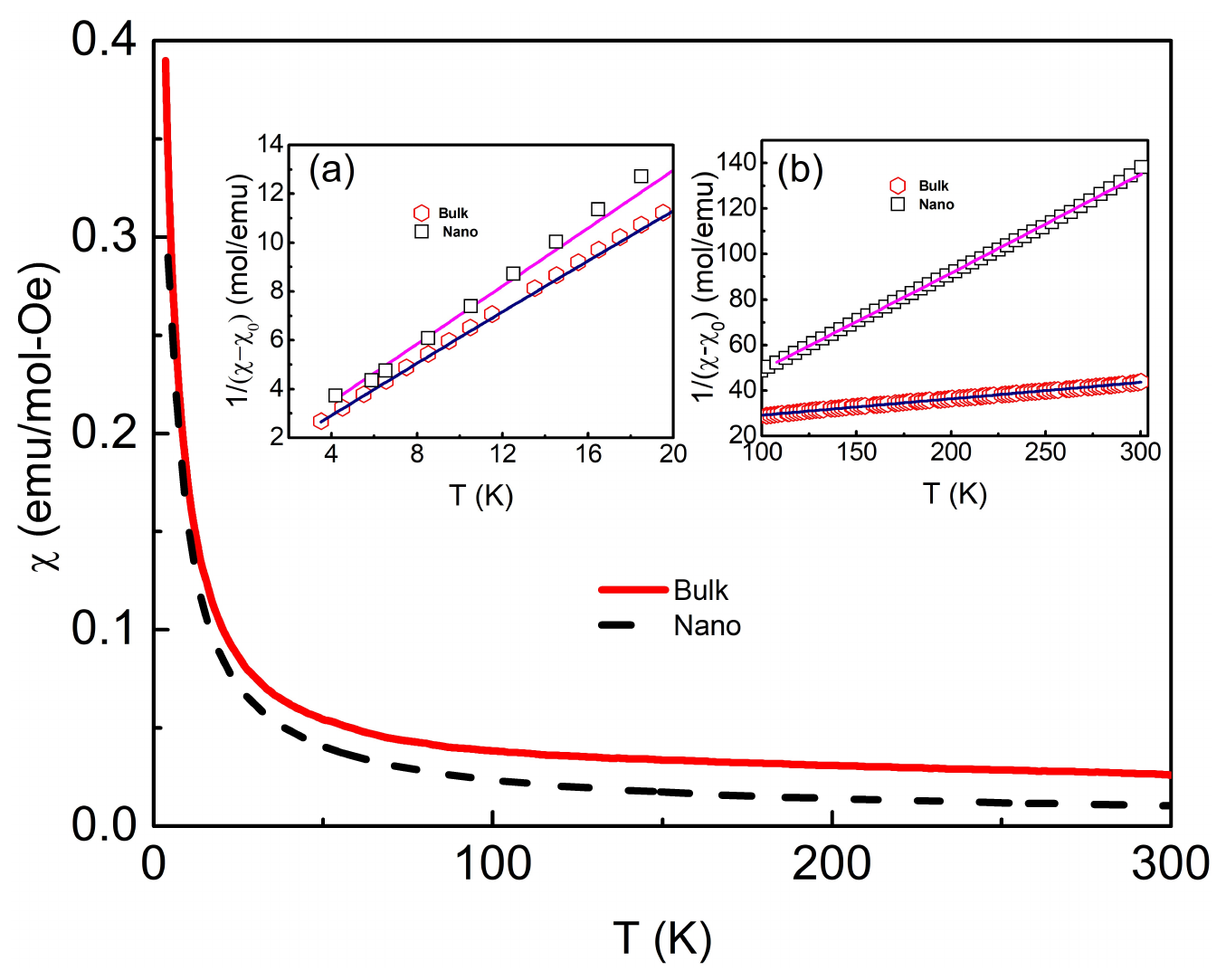}
\caption{Dc magnetic susceptibility of bulk and nanocrystalline PIO. Only FC data is shown as there is no bifurcation between FC and ZFC susceptibility. Inset a: Curie-Weiss plot at low temperature showing reduced $\theta_{CW}$ for both samples. Inset b: Curie-Weiss plot at high temperature for bulk and nanocrystalline PIO. No significant difference was observed in the high temperature $|\theta_{CW}|$ between the two samples.}\label{fig3}
\end{figure}

Fig.~\ref{fig1}e-g shows XPS spectra taken at room temperature for the bulk polycrystalline and nanocrystalline PIO.  To obtain a quantitative information about oxidation state, we fitted the photoemission peaks by XPS peakfit4.1 software using asymmetric Gauss-Lorentz sum profile using the protocol reported elsewhere \cite{Gamza}. For Pr - 3d spectra (Fig.~\ref{fig1}e), fitting suggests a majority of $Pr^{+3}$ charge state. The Pr - 3d spectra consists of main peaks $Pr - 3d_{5/2}$ and $Pr - 3d_{3/2}$ labeled as $f^{2}$  and the satellite peaks $f^{3}$. Here the main peaks appears due to masking of core hole by conduction electrons. Spectral character for Pr-3d is almost similar in bulk and nanocrystalline PIO with binding energy difference $\sim$ 20 eV among the main peaks. The spin-orbit coupling ($\sim3 eV$) splits the Ir 4f peak into low energy Ir 4$f_{7/2}$ and high energy Ir 4$f_{5/2}$, which is consistent with standard reported Ir 4f spectra \cite{Wertheim}. The de-convoluted XPS peaks of Ir 4f (Fig.~\ref{fig1}f) suggest presence of mixed oxidation state in both samples. The XPS peaks situated at 62 eV and 65.1 eV are attributed to $Ir^{+4}$ oxidation state of the 4$f_{7/2}$ and Ir 4$f_{5/2}$ respectively, while the corresponding binding energies of $Ir^{+5}$ peaks are at 64.2 eV and 67.8 eV for bulk polycrystalline PIO. The obtained binding energies of $Ir^{+4}$ and $Ir^{+5}$ peaks for nanocrystalline PIO are: $62.3$ eV and $65.4$ eV, $63.5$ eV and $66.8$ eV, respectively. The calculated peak area ratio of $Ir^{+5}$ : $Ir^{+4}$ for polycrystalline PIO is $0.25:1$ and for nanocrystalline PIO is $0.26:1$, respectively, showing negligible difference between the two. Thus the mixed oxidation state should not be responsible for differences in electrical conductivity in the two samples, contrary to what is observed in Y$_{2}$Ir$_{2}$O$_{7}$\cite{Dwivedi}. As shown in Fig.~\ref{fig1}(g), the two peaks are clearly resolved in O-1s spectra, labeled as $O_{HBE}$($\sim$ 531eV) and $O_{LBE}$ ($\sim $529 eV). The $O_{HBE}$ peak is possibly associated with the presence of O-$Ir^{+4}$ or O-$Pr^{+4}$ or both\cite{Nov}, while on the other hand, the $O_{LBE}$ peak could be attributed to $O^{2-}$ anion in the system.

Fig.~\ref{fig2}a shows temperature dependence of spin contribution to the resistivity ($\rho_{m}$) for bulk polycrystalline and nanocrystalline PIO. We calculated $\rho_{m}$ after subtracting $\rho_{min}$  from $\rho(T)$  at low temperature. We observe a shallow resistivity minima at $27$~K and $55$~K for bulk and nanocrystalline PIO, respectively, below which the resistivity increases upon further cooling. Although both samples show qualitatively similar behaviour, the enhancement of resistivity is significantly higher in nanocrystalline PIO ($\sim 21\%$) compared to the bulk polycrystalline sample ($\sim3\%$). Overall, the spin contribution to the resistivity follows generalized Hamann's expression incorporating potential scattering at each impurity site\cite{Hamann,Fischer}. The extracted values of $T_{K}$ are $7$~K and $18$~K for bulk and nanocrystalline PIO, respectively, suggesting that Kondo correlation is stronger in the latter. The residual resistivity were estimated by extrapolating Hamanns's fit to $T=0$ (given in Table.~\ref{T1}). Another well-known experimental signature of Kondo effect is the universal scaling of spin resistivity\cite{Schilling}. The reduced spin contribution to resistivity $\rho_{m}/\rho_{0}$ overlaps with each other when plotted as a function of $T/T_{K}$ at least down to $T_{K}$ (not shown in figure). In Fig.~\ref{fig2}b, we show the the partial suppression of upturn and the resulting negative magnetoresistance (MR) for both poly and nanocystalline PIO at constant magnetic field. At $T>>T_K$, the negative MR is negligibly small and independent of temperature and field for both samples. Interestingly, at low temperature, just above $T_K$, the negative MR increases sharply with decreasing temperature. Further, the negative MR increases with increasing field in this regime. For $T<<T_K$, at low field, the MR is proportional to $H^{2}$ for both the samples (Inset, Fig.~\ref{fig2}a).

The low temperature specific heat data of bulk polycrystalline and nanocrystalline PIO, as shown in Fig.~\ref{fig2}c, clearly proves existence of linear-T component. The linear coefficient $\gamma$ as well as the Debye temperature $\theta_D$ extracted from the $T^3$ contribution to the specific heat are listed in table~\ref{T1}. Normally, Kondo system shows large value of Sommerfeld coefficient $\gamma$ due to quasi-particle mass enhancement \cite{singh,sakurai}. The value of Sommerfeld coefficients for bulk polycrystalline and nanocrystalline sample are $235$ and $288$ times larger than in copper ($\gamma=0.695$ mJ/molK$^{2}$). The Sommerfeld coefficient for bulk single crystal PIO extracted from reported data~\cite{Takatsu} is almost half the $\gamma$ value for the bulk system in our case. Such enhancement of linear specific heat coefficient in PIO could be a consequence of Kondo screening. The renormalization factor for quasi-particle density of states due to 4f correlation in nanocrystalline PIO when compared to bulk polycrystalline PIO can be estimated by the ratio $\gamma_{nano}/ \gamma_{bulk}$, which turns out to be $\approx 1.2$. At low temperature, there is clear deviation from the Fermi liquid behaviour, which will be addressed elsewhere~\cite{Bikash}.
\begin{figure}
\includegraphics[width=\linewidth]{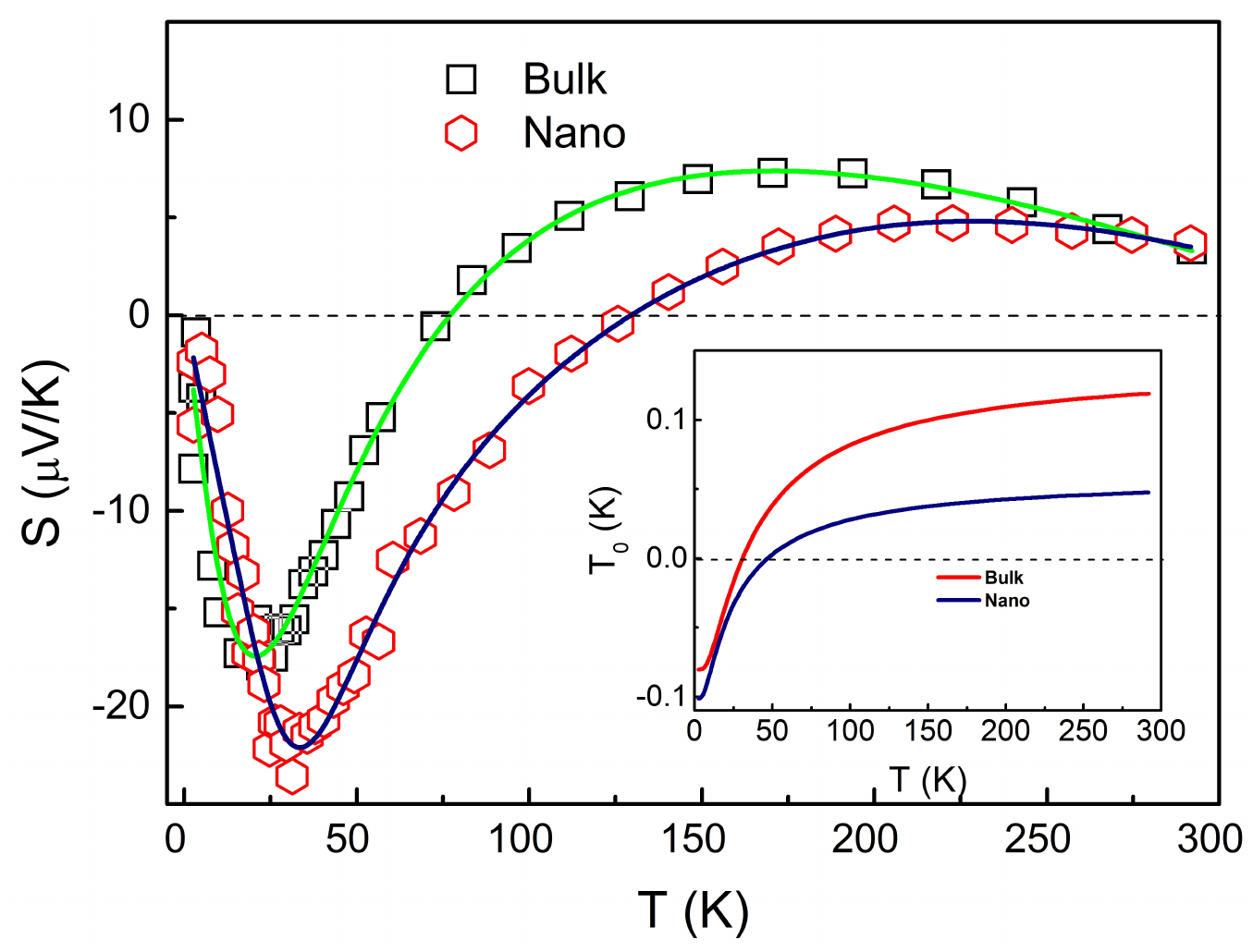}
\caption{The temperature dependence of thermopower ($S$) for bulk and nanocrystalline PIO showing sign inversion along with a positive maximum ($S_{max}$) at high temperature ($T_{max}$) and negative minimum ($S_{min}$) at low temperature ($T_{min}$) for both samples. The continuous lines are the fits using the phenomenological model (Eq.~\ref{eqn1}) discussed in the text. Inset: The fitting parameter $T_0$ (discussed in the text) for the two samples, which change sign with temperature, are also plotted.}\label{fig4}
\end{figure}

The temperature dependence of magnetic susceptibilities of bulk and nanocrystalline PIO are shown in Fig.~\ref{fig3}. We find absence of bifurcation in FC and ZFC $\chi(T)$ of both the samples (not shown in the figure) and $\chi(T)$ follows modified Curie-Weiss law, $\chi(T)=\chi_{0}+C/(T - \theta_{CW})$, where $\chi_{0}$ is the temperature independent susceptibility contributed by both $Pr^{+3}$ and $Ir^{+4}$, C is Curie constant and $\theta_{CW}$ is the Curie-Weiss temperature. \textcolor{blue}{ In general, the Curie-Weiss law is suitable to describe a paramagnetic insulator, it can also be applied for a metal having low carrier concentration in the non-degenerate limit ~\cite{Blundell}. In PIO however the  $Pr^{+3}4f$ shell magnetic moments are far less likely to be interacting with one another owing to the high degree of localization. The Pauli paramagnetic contribution from $Ir5d$ itinerant electrons and Van Vleck term arising from the narrow energy separation ($\sim 160K$) among non-Kramers doublet ground state and the excited Kramers doublet of $Pr^{+3} 4f^{2}$, are captured in $\chi_{0}$ and hence the justification for the use of Curie-Weiss law for metallic PIO.} The high $T$ (above $100$ K) value of Curie-Weiss temperature $\theta_{CW}$ are $-19.1$ K and $-18.9$ K for bulk and nanocrystalline PIO, respectively. If the AF interaction, appearing due to RKKY coupling~\cite{Lee,Ikeda}, sets the scale of Kondo coupling, the value of T$_K$ in nanocrystalline sample should be marginally reduced or remain unchanged, which is clearly not the case here. The estimated value of high T effective magnetic moment $\mu_{eff}$ for bulk and nanocrystalline PIO are 3.80 $\mu_{B}$/Pr and 3.75 $\mu_{B}$/Pr, respectively, very close to the magnetic moment of a free $Pr^{+3}$ ($\mu _{eff}=3.58\mu_{B}$) ion. The corresponding reported values of $\theta_{CW}$ and $\mu_{eff}$ for bulk single crystals are $-22.8$ K and $3.3\mu_B$/Pr, respectively~\cite{Machida}. Though $Ir^{+4}$ has five electrons in 5d orbital, the crystal electric field due to local oxygen octahedral environment substantially lowers its effective magnetic moment from the expected Hund's rule value. As a result of spin-orbit coupling, four electrons completely fill energetically lower quartet $(J_{eff}=3/2)$ of the $t_{2g}$ level while the single unpaired electron in the $J_{eff}=1/2$ state contributes to the magnetic moment of $Ir^{+4}$, giving $\mu_{Ir}=\frac{1}{3}\mu_{B}$~\cite{Shapiro}. The resultant theoretical effective moment of PIO ($\approx\sqrt{2\mu_{Pr}^{2}+2\mu_{Ir}^{2}}$) with and without Ir moment is 5.08$\mu_{B}$/f.u. and 5.06$\mu_{B}$/f.u., respectively. Thus, the contribution of $Ir^{+4}$ to the total effective moment can be neglected for all practical purposes. \textcolor{blue}{ In contrast, for $Pr^{+3}$ ion, the 4f electrons are less extended from the nucleus and well shield from the surrounding crystal electric field by the 5s and 5p shells, resulting the orbital moment remains unquenched ~\cite{Blundell}. However, the overlapping of the radial wave functions bring the highly localized $Pr4f$ electrons (small but nonvanishing wavefunction exist beyond 5s and 5p shell) and $Ir5d$ itinerent  electrons into a quantum mechanical contact and giving rise  an antiferromagnetic coupling among them ~\cite{Coleman} (The parameter W in Eq.~\ref{eqn1} and Eq.~\ref{eqn2} represents the hybridization strength between $4f$ band and conduction band.).} 

The ground state of a Kondo system is a singlet with total spin $S=0$, while above $T_K$, the local moment is unscreened with $S\neq0$. At low temperature, the conduction electrons screen the local $Pr^{+3}$ moment and the efficiency of the screening increases with decreasing $T$, which, in turn, reduces the effective moment. We observe that, below $20$~K, the value of $\mu_{eff}$ as well as $\theta_{CW}$ are diminished considerably (insets a, b, Fig.~\ref{fig3}), similar to the case for single crystalline PIO~\cite{Nakatsuji}. The low temperature effective magnetic moment and Weiss temperature for bulk and nanocrystalline PIO are (2.69 $\mu _{B}$, -1.4K) and (2.67 $\mu _{B}$/Pr, -1.9K), respectively, suggesting \textcolor{blue}{ a possible} screening of magnetic moments below $T_K$. The value of $\mu_{eff}$ is also calculated from isothermal magnetization curve (not shown in the figure) by fitting with Brillouin function, which gives 2.52$\mu_{B}$/Pr and 2.46$\mu_{B}$/Pr for bulk and nanocrystalline sample, respectively, which are in fair agreement with the values obtained from $\chi(T)$.  \textcolor{blue}{Additionally, the ground state of the $4f^{2}$ configuration of $Pr^{+3}$ ion under the cubic crystalline electric field (CEF) is assumed to be the non-Kramers doublet with active quadrupoles ~\cite{Cox1}. In $Pr$-based intermetallic systems however the prerequisite for the quadrupole Kondo effect can be fulfilled as proposed by Cox ~\cite{Cox1}. In the quadrupolar Kondo effect, local quadrupole is coupled by two-channel conduction bands and exhibits logarithmic temperature dependence of susceptibility.  Therefore, the appearance of the temperature dependence of the effective magnetic moment due to the coexistence of magnetic and quadrupolar Kondo effect or an antiferroquadrupolar order could not be ruled out either ~\cite{Cox1,Onimaru,Ni}. }   

\begin{figure}
\includegraphics[width=\linewidth]{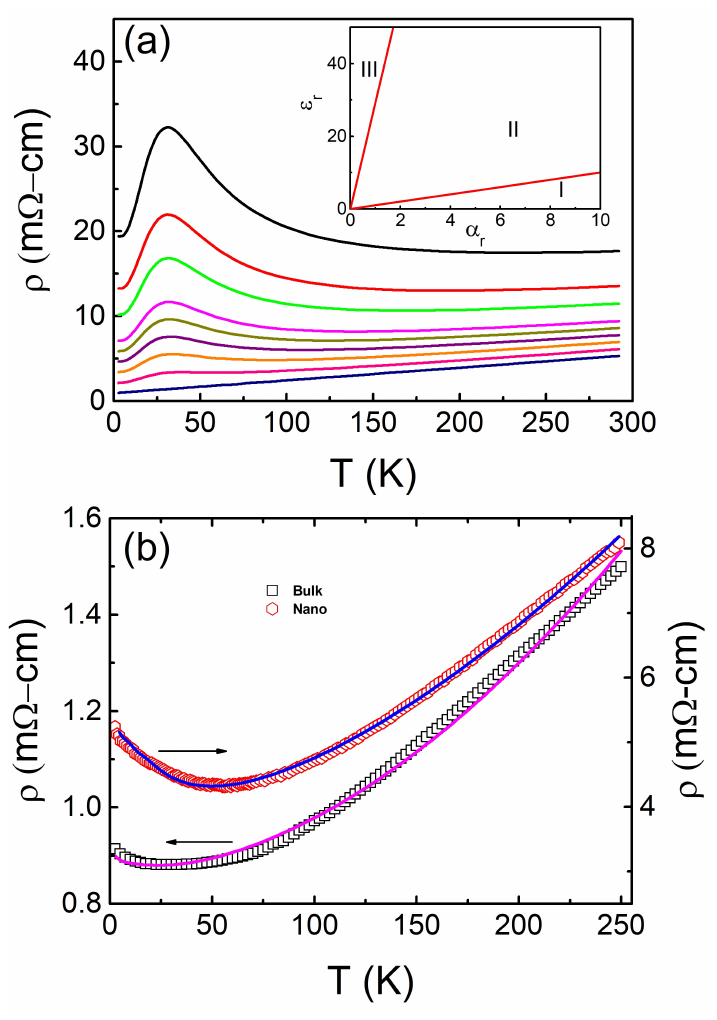}
\caption{(a) The simulated temperature dependence of resistivity using Eq.~\ref{eqn2}, excluding the $T^2$ term. The parameters $a$ and $E$ were varied while using the fixed parameters $T_m$ and $T_f$ obtained from fitting of thermopower data using Eq.~\ref{eqn1}. Inset: The constants $a$ and $E$ are expressed as dimensionless parameters $\alpha_r$ and $\epsilon_r$, respectively. The corresponding different transport regimes as predicted by Eq.~\ref{eqn2} (excluding the quadratic term), depending on the ratio of $\epsilon_r$ and $\alpha_r$ are labeled as I, II, III (discussed in the text). The diagram shows that resistivity minimum without the coherence maximum cannot be observed in absence of the quadratic term in Eq.~\ref{eqn2}. (b) The temperature dependence of resistivity for bulk and nanocrystalline sample fitted using Eq.~\ref{eqn2}.}\label{fig5}
\end{figure}

\begin{table*}
\begin{center}
\begin{tabular}{c c c c c c c c c c c}
\hline
 & $\rho_0$ & $T_K$ & $|\theta_{CW}|$ & $\gamma$   & $\beta$  & $\Theta_{D}$ & $T_{max}$ & $S_{max}$  & $S_{min}$ & $T_{min}$\\
 & ($m\Omega cm$) & ($K$) & ($K$) &(mJ/mol$K^{2}$) &(mJ/mol$K^{4}$) & ($K$) & ($K$) & ($\mu V/K$)  & $(\mu V/K$) & ($K$) \\
 \hline
 Bulk & 0.9 & 7 & 19.1 & 165 &  0.258  &  436  & 173  & 7.3  & -18 & 16 \\
 Nano & 5.2 & 18 & 18.9 & 200 &  0.407  &  375  & 230  & 4.6  & -24 & 30 \\
  \hline
\end{tabular}
\end{center}
\caption{Comparison of important physical parameters extracted from transport and thermodynamic measurements.\label{T1}}
\end{table*}

Before moving into thermopower data and further analysis of resistivity, which is the main content of the present article, let us summarize what we have discussed so far. The spin contribution to the resistivity, for both bulk polycrystalline and nano-crystalline PIO, show universal scaling behaviour. The Kondo temperature obtained from Hamann's fit suggests that Kondo coupling is enhanced in nano-crystalline PIO compared to bulk poly-crystalline sample. By using modified Curie-Weiss law to fit magnetic susceptibility, we extract effective magnetic moment ($\mu _{eff}$) at low and high temperature in both samples. We observe that $\mu_{eff}$ decreases substantially at low temperature ($T<<T_K$) compared to the same at $T>>T_K$. Moreover, from low temperature specific heat ($C_{P}$) measurement, we determine the electronic and lattice contribution to $C_{P}$. An enhanced linear specific heat ($\gamma T$) is observed for both samples, indicating re-normalization of thermodynamic mass of quasi-particles. The most interesting observation is that the Kondo coupling scale $T_K$ has got very little correlation with the RKKY coupling scale represented by $|\theta_{CW}|$ at high temperature.

\begin{table*}
\begin{center}
\begin{tabular}{c c c c c c c c c c}
\hline
 & $C_1$  & $C_2$  & $A$ & $B$ & $T_m$ & $T_f$ & $a$  & $b$  & $E$  \\
 & ($\mu V/K^2$) & ($\mu V/K$) &  ($K$)   & ($K$)    &  ($K$)     &   ($K$) & ($m\Omega cm/K$) &  ($m\Omega cm/K^2$) & ($m\Omega cm/K$) \\
 Bulk  & -0.5 &  0.10 & ~-0.08~ & ~0.22~ & ~31.3~ & ~0.10~ & 5.4$\times$10$^{-4}$ & 9.1$\times$10$^{-6}$ & 0.086\\
 Nano  & -0.4 &  0.04 &  ~-0.10~ & ~0.16~ & ~21.6~ & ~0.05~ & ~0.01~ & 39.6$\times$10$^{-6}$ & 0.24 \\
\hline
\end{tabular}
\end{center}
\caption{Fitting parameters in the phenomenological model describing temperature dependence of thermopower (equation.~\ref{eqn1}) and temperature dependence of resistivity (equation.~\ref{eqn2}). \label{T2}}
\end{table*}

In Fig.~\ref{fig4}, we plot the temperature dependence of thermopower ($S$) for bulk and nanocrystalline PIO. Both samples show qualitatively similar behavior such as, a broad positive maximum $S_{max}$ at $T_{max}$ followed by negative minimum $S_{min}$ at $T_{min}$ at lower temperature. The values of ($S_{max}, T_{max}$) and ($S_{min}, T_{min}$) for bulk and nanocrystalline PIO are given in table~\ref{T1}. In the present case, the temperature dependence of thermopower have similar feature to the low $T_{K}(\sim10K$) Ce based Kondo systems such as $CeAl_{2}$, $CeCu_{2}Si_{2}$ \cite{ray, brandt}. The maximum in thermopower at high temperature $S_{max}$ arises due to crystalline field effects in a Kondo lattice system~\cite{bhattacharjee}. In this case, the maximum for the bulk polycrystalline sample is at $173$ K which is very close to crystal field splitting $\Delta_{CF}$ between the Pr$^{3+}$ non-Kramers doublet ground state and the Kramers doublet excited state reported earlier~\cite{Nakatsuji}. The maximum becomes shallow and shifts to higher temperature ($\sim 230$ K) for the nanocrystalline sample. The thermopower turns negative well above T$_K$ for both samples. A negative thermopower is observed when hole scattering becomes dominant at low temperatures, which is consistent with strong charge fluctuations in a two channel Kondo system~\cite{cox}. Moreover, the large negative thermopower shows a minimum at $T_{min}=16$~K for the bulk system which is very close to the antiferromagnetic interaction $|\theta_{CW}|\sim 19$~K attributed to inter-site RKKY coupling. This seems to suggest a `two impurity' Kondo scattering dominated by RKKY interaction. However, the same minimum for the nano-crystalline sample occurs at $T_{min}=32$~K although corresponding $|\theta_{CW}|$ remains unchanged. The higher value of $S_{min}$ and $T_{min}$ for nanocrystalline PIO compared to the bulk sample is consistent with the enhanced $T_K$ in the former. The ratio of $S_{min}$ in the two samples is $1.3$, which is in fair agreement with the value obtained from specific heat measurement ($\gamma_{nano}/\gamma_{bulk}\sim$1.2). The observation of enhanced $T_{max}$, $T_{min}$ and $T_K$ in the nano-crystalline sample with negligible change in $\theta_{CW}$, strongly supports the possibility of a mechanism other than the single channel inter-site Kondo effect. The difference of $T_K$ in the two samples could be attributed to the disorder within the context of a single impurity Kondo problem, where $T_K$ is, \emph{inter alia}, related to the disorder induced local density of states fluctuation~\cite{Dobro}. Surprisingly, it is the bulk polycrystalline sample which is more sensitive to disorder compared to the nanocrystalline sample as $T_K$ is relatively suppressed in the former.

The temperature dependence of thermopower can be described using a phenomenological model of a Kondo lattice, including the effect of crystal field excitation~\cite{ray}.
\begin{equation}
S=C_1 T + \frac{C_2 T T_0}{{T_0}^2 + W^2}\label{eqn1}
\end{equation}
Here, $W=T_f \exp(-T_f /T)$ and $T_0 = A + B \exp(-T_m / T)$, and $C_1$, $C_2$, $T_f$, $T_m$, $A$, $B$ are the constant parameters. The parameters extracted from the fitting (as shown in Fig.~\ref{fig4}), are given in Table.~\ref{T2}. As expected, $T_0$ is temperature dependent and changes sign at approximately the same temperature as the corresponding $S$ (inset, Fig.~\ref{fig4}). The temperature dependence of the parameter $T_0$ along with sign inversion suggest that the system is more likely heavy fermionic (HV) rather than mixed valent (MV)~\cite{ray}. We find that $T_m$, which is an approximate measure of effective separation between CF split levels due to level broadening, decreases in the nanocrystalline sample compared to the bulk. This is expected as the level broadening is higher in the nanocrystalline sample compared to the bulk due to the enhanced $T_K$ in the former. The question arises then as to what causes the thermopower peak to shift towards higher temperature for nanocrystalline sample. We do not expect any significant increase in crystal field splitting due to the marginal reduction in unit cell volume. The high temperature peak develops when the two terms in Eq.~\ref{eqn1} becomes comparable, which implies that the peak position is dependent on several competing factors. Although there is a decrease of $T_m$ (due to the crystal field level broadening) and enhancement of $C_1/C_2$ (or relative increase of nonmagnetic scattering contribution), which should, in principle, lead to lowering of peak position, enhancement of $A/B$ (a measure of the ratio of the proximity of the hybridized band to the Fermi level in the limit $T=0$ and $T=\infty$, respectively) and reduction of $T_f$ (a measure of the bandwidth) in the nanocrystalline sample lead to the increase of $T_{max}$ in the same.

Using the parameters obtained from thermopower data, one can also simulate the temperature dependence of resistivity as follows.
\begin{equation}
\rho=\rho_0 + aT + bT^2 + E \frac{W}{{T_0}^2+W^2} \label{eqn2}
\end{equation}
Here $\rho_0$ is the residual resistivity, $a$ is constant term associated with nonmagnetic contribution to resistivity and $E$ is another constant associated with the magnetic contribution. We have added a $bT^2$ term, where $b$ is another constant, for reasons to be discussed shortly hereafter.

In principle, the expression in Eq.~\ref{eqn2} can reproduce the observed resistivity minimum and the saturation at low temperature even without the $T^2$ term. However once we fix the parameters obtained from the thermopower fit such as $T_0$ and $W$, we find that the resistivity minimum is always accompanied by a coherence maximum with a drop in resistivity at lower temperature. In order to demonstrate this, we plot different possible scenarios in absence of $T^2$ term, taking into account the variation of $a$ and $E$ under the constraints that other important parameters are fixed by the thermopower fit (Fig.~\ref{fig5}a). The coherence maximum occurs around the temperature $\sim T_m$ and is largely insensitive to the variation in $a$ and $E$. It can only be suppressed together with the resistive minimum leading to metallic resistivity (Fig.~\ref{fig5}a).

We construct a diagram describing different possible regimes of electrical conduction within the experimental temperature range, and under the constraint that the parameters extracted from fitting of the thermopower data remain unaltered (Inset, Fig.~\ref{fig5}a). The $3$ distinguishable regimes depending on the relative contribution of $a$ and $E$ are as follows: $1)$ Region I: broadly metallic with no resistive peak ($T^\ast$) and no resistive minimum ($T_{min}$); $2)$ Region II: coherent Kondo lattice regime with observation of both $T^\ast$ and $T_{min}$; Region III: broadly insulator like temperature dependence with no $T_{min}$ but appearance of resistive peak at $T^\ast$. We express $a$ and $E$ as dimensionless variables $\alpha_r$ and $\epsilon_r$ (normalized by any point in the two-parameter space), respectively. When $\frac{\epsilon_r}{\alpha_r} <1$, i.e, when phonon contribution dominates over Kondo scattering contribution, $\rho(T)$ shows completely metallic behaviour (region I). In region II, Kondo contribution is comparable but moderately higher than phonon contribution ($1\leq\frac{\epsilon_r}{\alpha_r}\leq29)$. In this regime resistivity minima is always accompanied by a precipitous drop at lower temperature. When Kondo contribution dominates over the nonmagnetic phonon contribution ($\frac{\epsilon_r}{\alpha_r}>29$), $\rho(T)$ shows only the resistive peak and the minimum disappears (region III).

Clearly, while the temperature dependence of thermopower can be explained within the framework of the phenomenological Kondo lattice model given by Eq.~\ref{eqn1}, the presence of the shallow resistance minimum without any coherence maximum in the experimental resistivity data does not match with the model prediction in Eq.~\ref{eqn2}, in absence of the quadratic term. Moreover, the phenomenological model in itself does not reproduce the $T^2$ term in the low temperature limit either~\cite{ray} (which is a characteristic of coherent Kondo lattice). This provides adequate justification for including the $T^2$ term in Eq.~\ref{eqn2}, similar to a two fluid description incorporating a coherent contribution to resistivity. Once the $T^2$ term is considered, the temperature dependence of resistivity for both bulk and nanocrystalline sample in a wide temperature regime is well described by Eq.~\ref{eqn2}, as shown by the fitting in Fig.~\ref{fig5}b.

To conclude, we have provided experimental evidence of Kondo origin of resistivity minima in both bulk and nanocrystalline PIO. The Kondo effect was confirmed, in addition to the usual transport signatures, by the diminished value of $\mu_{eff}$ and $\theta_{CW}$ in low temperature susceptibility, enhanced linear coefficient in specific heat measurement as well as negative maximum in the thermopower at low temperature. The value of $T_{K}$ is substantially enhanced in nanocrystalline PIO while the high temperature $\theta_{CW}$ remains unchanged compared to bulk polycrystalline PIO. Moreover, the temperature dependence of resistivity and thermopower can be fully reproduced by the phenomenological model describing crystal field excitation in a Kondo lattice system.

We acknowledge Department of Science and Technology (DST), government of India for financial support.

\end{document}